\documentclass[showpacs,preprintnumbers,amsmath,amssymb,twocolumn]{revtex4}%
\usepackage{doi}%<----------
\usepackage{hyperref}
\hypersetup{
  colorlinks=true,        % false: boxed links; true: colored links
  linkcolor=blue,         % color of internal links
  citecolor=cyan,         % color of links to bibliography
}

\usepackage{graphicx}% Include figure files
\usepackage{dcolumn}% Align table columns on decimal point
\usepackage{bm}% bold math
\usepackage{color}

\begin{document}
\title{Optical properties of black hole in the presence of plasma: shadow}
\author{Farruh Atamurotov$^{1,2}$}
   \email{farruh@astrin.uz}
\author{Bobomurat Ahmedov$^{1,3}$}
 \email{ahmedov@astrin.uz}
\author{Ahmadjon Abdujabbarov$^{1,3}$}
 \email{ahmadjon@astrin.uz}

\affiliation{%
$^{1}$ Institute of Nuclear Physics, Ulughbek, Tashkent 100214,
Uzbekistan\\
$^{2}$ Inha University in Tashkent, Tashkent 100170, Uzbekistan\\
$^{3}$Ulugh Beg Astronomical Institute, Astronomicheskaya 33,
Tashkent 100052, Uzbekistan }
\begin{abstract}
We have studied photon motion around axially symmetric rotating Kerr black hole in the presence of  plasma with radial power-law density.
It is shown that in the presence of plasma the observed shape and size of shadow changes depending on i) plasma parameters, ii) black hole spin
and iii) inclination angle between observer plane and axis of rotation of black hole. In order to extract pure effect of plasma influence on
black hole image the particular case of the Schwarzschild black hole has also been investigated and it has been shown
that i) the photon sphere around the spherical symmetric black hole is left unchanged under the plasma influence, ii) however
the Schwarzschild black hole shadow size in plasma is reduced due to the refraction of the electromagnetic radiation
in plasma environment of black hole.
The study of the energy emission from the black hole in plasma shows that in the presence of plasma the maximal energy emission rate from the black hole decreases.
\end{abstract}
\maketitle
\section{Introduction}
 The study of astrophysical processes in plasma medium surrounding black hole becomes  very interesting and important
  due to the evidence for the presence of black holes at the centres of the galaxies ~\cite{Eatough13,Falcke00,Falcke01}.
 For example, the gravitational lensing in inhomogeneous and homogeneous plasma around black holes has been
 recently studied in ~\cite{Bisnovatyi2010,Tsupko12,Morozova13,Perlick15,Er14} as extension of vacuum studies (see, e.g. ~\cite{Schee09,Schee15}).

 From the literature it is known that the black hole shadow is appeared by the gravitational lensing effect, see, e.g. ~\cite{Amarilla12,Tsukamoto14,Atamurotov13b,Falcke00}.
 If black hole is placed between a bright source and far observer, dark zone is created in the source image by photon fall inside black hole which is
  commonly called shadow of black hole.  Recently, this effect is investigated by many authors
   for the different black holes (see, e.g. ~\cite{Hioki09,Grenzebach2014,Amarilla10,Atamurotov13}).
   The silhouette shape of an extremely rotating black hole has been investigated by Bardeen ~\cite{Bardeen73}.
   Our previous studies on the shadow of black hole are related to the non-Kerr~\cite{Atamurotov13b},
   Ho\v{r}ava-Lifshitz~\cite{Atamurotov13}, Kerr-Taub-NUT~\cite{Abdujabbarov13c} and Myers-Perry~\cite{Papnoi14} black holes.
    A new coordinate-independent formalism for characterization of a black-hole shadow has been recently developed in ~\cite{Abdujabbarov15}.

   Shape of black hole is determined through boundary of the shadow which can be  studied by application of the null geodesic equations.
   The presence of plasma in the vicinity of black holes changes the equations of motion of photons
   which may lead to the modification of black hole shadow by the influence of plasma.
In this paper our main goal is to consider silhouette of shadow of axially symmetric black hole using the equations of motion for photons
in plasma with radial power-law density.  We would like to underline that very recently, influence of a non-magnetized cold plasma 
with the radially dependent density 
to black hole shadow has been studied in~\cite{Perlick15} using the different alternate approach. In addition~\citet{Rogers15} has studied the photon motion around black hole surrounded by plasma.

The paper is arranged as following. In Sect.~\ref{geodesics}, we consider the
 equations of motion of photons around axially symmetric black hole in the presence of plasma. In
Sect.~\ref{bh-shadow} we study the shadow of the axial-symmetric black hole in the presence
of  plasma. As particular case in subsections \ref{nonrot} and \ref{emission} we study the shadow and the energy emission from the sphericaly symmetric black hole.
Finally, in Sect.~\ref{conclusion} we briefly summarize the results found.

Throughout the paper, we use a system of geometric units in which $G = 1 = c$. Greek indices run from $0$ to $3$.

\section{Photon motion around the black hole in the presence of plasma}
\label{geodesics}

The rotating black hole is described by the space-time metric, which in
the standard Boyer-Lindquist coordinates, can be written in the form
\begin{equation}
ds^2 = g_{\alpha \beta}dx^\alpha dx^\beta\ , \label{metric}
\end{equation}
with ~\cite{chandra98}
\begin{eqnarray}
g_{00}&=&-\left(1-\frac{2 M r}{\Sigma}\right)\ , \nonumber \\
 g_{11}&=&\frac{\Sigma}{\Delta}\ , \nonumber \\
g_{22}&=&\Sigma\ , \nonumber \\
 g_{33}&=&\left[(r^2+a^2)+\frac{2 a^2 M r \sin^2
\theta}{\Sigma}\right]\sin^2
\theta\ , \nonumber \\
 g_{03}&=&-\frac{2 M a r \sin^2 \theta}{\Sigma}\ ,\label{1}
\end{eqnarray}
\begin{equation}
\Delta = r^2 + a^2 -2 M r , \;\;\;\;\; \Sigma=r^2+a^2 \cos^2\theta
\ , \nonumber \label{p2}
\end{equation}
where as usual $M$ and $a$ are the total mass and the spin parameter of
the black hole.

In this paper we will consider the plasma surrounding the central axially symmetric black hole. The refraction index of the plasma will be $n=n(x^{i}, \omega)$ where the photon frequency measured by observer with velocity $u^\alpha $ is $\omega$. In this case the effective energy of photon has the form  $\hbar \omega= - p_\alpha  u^\alpha$. The refraction index of the plasma as a function of the photon four-momentum has been obtained in~\citep{Synge60} and has the following form:
\begin{equation}
n^2=1+\frac{p_\alpha p^\alpha}{\left( p_\beta u^\beta \right)^2} ,
\end{equation}
and for the vacuum case one has the relation $n=1$. The Hamiltonian for the photon around an arbitrary black hole surrounded by plasma has the following form
\begin{equation}
H(x^\alpha, p_\alpha)=\frac{1}{2}\left[ g^{\alpha \beta} p_\alpha p_\beta + (n^2-1)\left( p_\beta u^\beta \right)^2 \right]=0.
\label{generalHamiltonian}
\end{equation}

Following to the derivation of a gravitational
redshift discussed in~\cite{Rezzolla04} we will assume that the spacetime
stationarity allows existence
of a timelike Killing vector $\xi^\alpha$ obeying to the Killing equations
\begin{equation}
\label{killing-eq}
\xi_{\alpha ;\beta}+\xi_{\beta;\alpha}=0\ .
\end{equation}

Then one can introduce two frequencies of electromagnetic waves using
null wave-vector $k^\alpha$ the first one is the
frequency measured by an observer with four-velocity $u^{\alpha}$ and
defined as
\begin{equation}
\label{freq_ob}
\omega \equiv -k^\alpha u_\alpha \ ,
\end{equation}
while the second one is the frequency associated with the timelike
Killing vector $\xi^{\alpha}$ and defined as
\begin{equation}
\label{freq_kil}
\omega_\xi \equiv -k^\alpha\xi_{\alpha} \ .
\end{equation}

    The frequency  (\ref{freq_ob})
depends on the observer chosen and is therefore a function of position,
while the frequency
(\ref{freq_kil}) is a conserved quantity that remains unchanged along the
trajectory followed by the electromagnetic wave.
One can apply  this property to measure how the frequency
changes with the radial position and is redshifted in the spacetime.
Assume the Killing vector to have components
\begin{equation}
\label{killing}
\xi^{\alpha}\equiv \bigg(1,0,0,0\bigg) \ ; \qquad
        \xi_{\alpha}\equiv g_{00} \bigg(- 1,0,0,0 \bigg) \ ,
\end{equation}
so that $\omega_{\xi}=k_0=$const. The frequency of an electromagnetic
wave emitted at radial position $r$  and measured by an observer
with four-velocity $u^{\alpha}\{1/\sqrt{-g_{00}},0,0,0\}$ parallel to $\xi^{\alpha}$ (i.e. a static
observer) will be governed by the following equation
\begin{equation}
\label{rs}
\sqrt{-g_{00}}\omega(r)= \omega_\xi={\rm const} \ .
\end{equation}
 One may introduce a specific form for the plasma frequency for analytic processing,
 assuming that the refractive index has the general form
\begin{equation}
n^2=1- \frac{\omega_e^2}{\omega^2},
\label{nFreq}
\end{equation}
where $\omega_e$ is usually called plasma frequency.
Now
using the Hamilton-Jacobi equation which defines the equation of motion of the photons for a given
space-time geometry ~\cite{Synge60,Rogers15,Bisnovatyi2010}:
\begin{equation}
\frac{\partial S}{\partial
\sigma}=-\frac{1}{2}\Big[g^{\alpha\beta}p_{\alpha}p_{\beta}-(n^2-1)(p_{0}
\sqrt{-g^{00}})^{2}\Big]\ , \label{p3}
\end{equation}
where $p_{\alpha}=\partial S/\partial x^\alpha$.
Using method of separation of variables the Jacobi action S can be written as
\cite{chandra98,Atamurotov13b}:
\begin{equation}
S=\frac{1}{2}m^2 \sigma - {\cal E} t + {\cal L} \phi +
S_{r}(r)+S_{\theta}(\theta)\ , \label{p4}
\end{equation}
where  $\cal L$, $\cal E$ are conservative quantities as angular
momentum and energy of the test particles.

For trajectories of the photons we have the following set of the equations:
\begin{eqnarray}
\Sigma\frac{dt}{d\sigma}&=&a ({\cal L} - n^2 {\cal E} a
\sin^2\theta)\nonumber\\&&+ \frac{r^2+a^2}{\Delta}\left[(r^2+a^2)n^2 {\cal
E} -a {\cal L} \right], \label{teqn}
\\
\Sigma\frac{d\phi}{d\sigma}&=&\left(\frac{{\cal L}}{\sin^2\theta}
-a  {\cal E}\right)+\frac{a}{\Delta}\left[(r^2+a^2) {\cal E}
-a {\cal L} \right], \label{pheqn}
\\
\Sigma\frac{dr}{d\sigma}&=&\sqrt{\mathcal{R}}, \label{reqn}
\\
\Sigma\frac{d\theta}{d\sigma}&=&\sqrt{\Theta}, \label{theteqn}
\end{eqnarray}
can be derived  from the Hamilton-Jacobi equation,
where the functions $\mathcal{R}(r)$ and $\Theta(\theta)$ are
introduced as
\begin{eqnarray}
\mathcal{R}&=&\left[(r^2+a^2) {\cal E} -a {\cal L}
\right]^2+(r^2+a^2)^2(n^2-1){\cal E}^2 \nonumber \\
&&
-\Delta\left[\mathcal{K}+({\cal L} -a {\cal E})^2\right]\ , \label{9}
\\
\Theta&=&\mathcal{K}+\cos^2\theta\left(a^2  {{\cal
E}^2}-\frac{{\cal L}^2}{\sin^2\theta}\right) \nonumber\\&& -(n^2-1) a^2 {\cal E}^2 \sin^2\theta\ , \label{10}
\end{eqnarray}
and the Carter constant as $ \mathcal{K} $.

For calculation examples one needs the analytical expression of the llama frequency $\omega_e$ which for the electron plasma has the following form
\begin{equation}
\omega_e^2=\frac{4 \pi e^2 N(r)}{m_e}
\label{plasmaFreqDef}
\end{equation}
where $e$ and $m_e$ are the electron charge and mass respectively, and $N(r)$ is the the plasma number density. Following the work by \citet{Rogers15} here we consider  a radial power-law density
\begin{equation}
N(r)=\frac{N_0}{r^h},
\label{powerLawDensity}
\end{equation}
where $h \geq 0$, such that
\begin{equation}
\omega_e^2=\frac{k}{r^{h}}.
\label{omegaN}
\end{equation}
As an example here we get the value for power $h$ as 1~\cite{Rogers15}. For this value we plot the radial dependence of the effective potential $V_{\rm eff}$ of the radial motion of the photons defined as
\begin{eqnarray}
\left(\frac{dr}{d\sigma}\right)^2+V_{\rm eff}=1\ .
\end{eqnarray}
\begin{figure*}[t!]
\begin{center}
\includegraphics[width=0.31\linewidth]{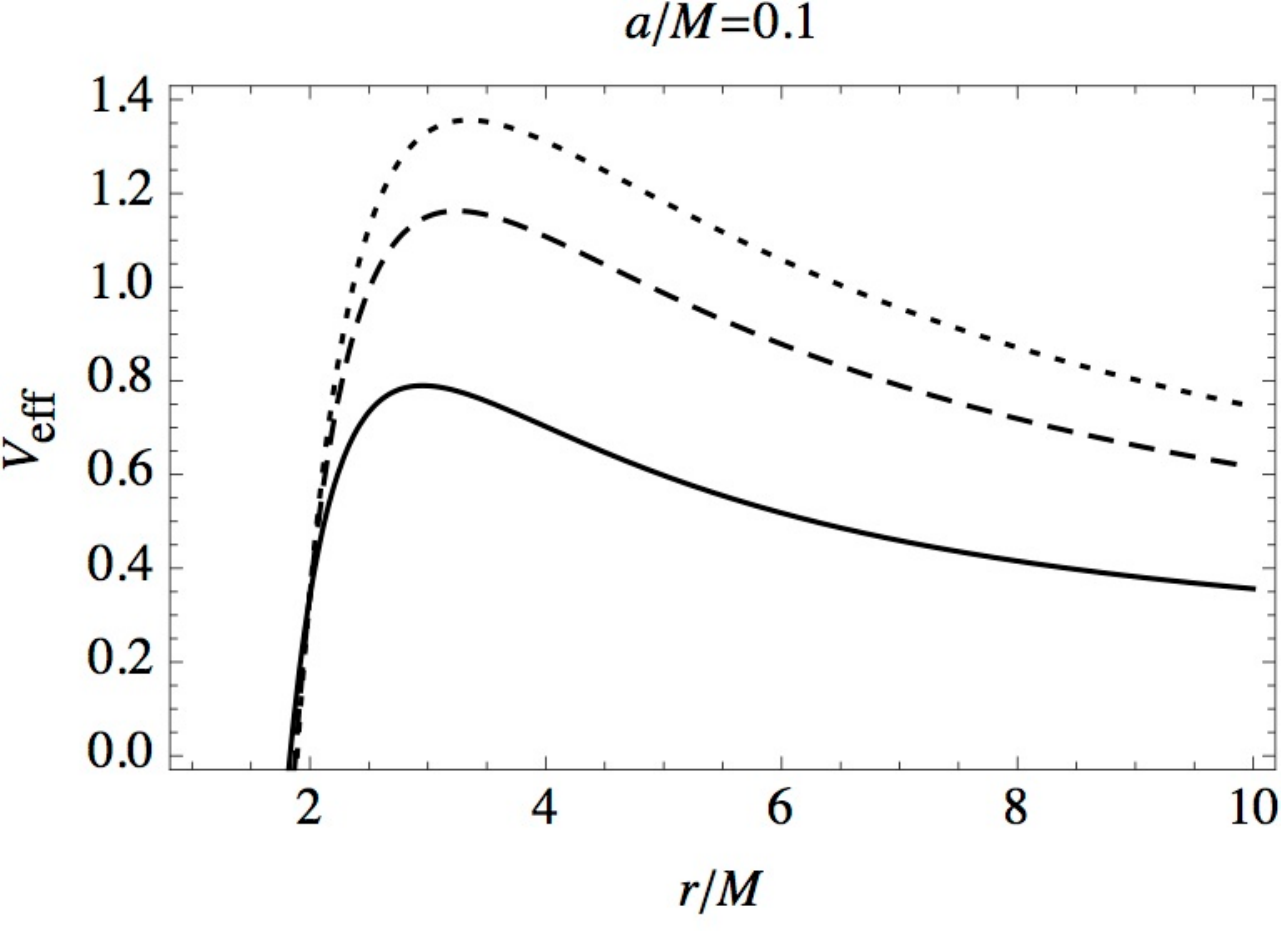}
\includegraphics[width=0.31\linewidth]{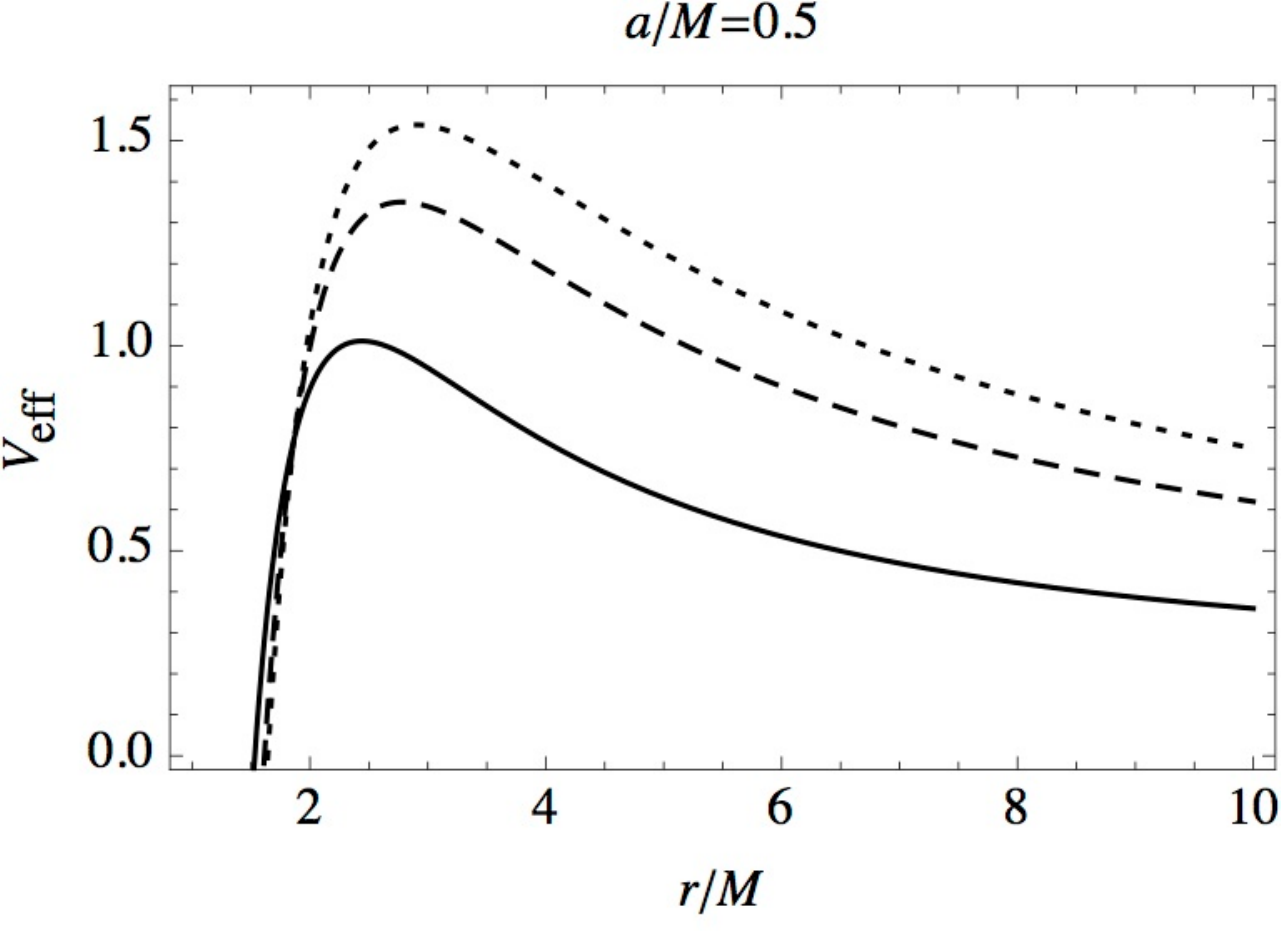}
\includegraphics[width=0.31\linewidth]{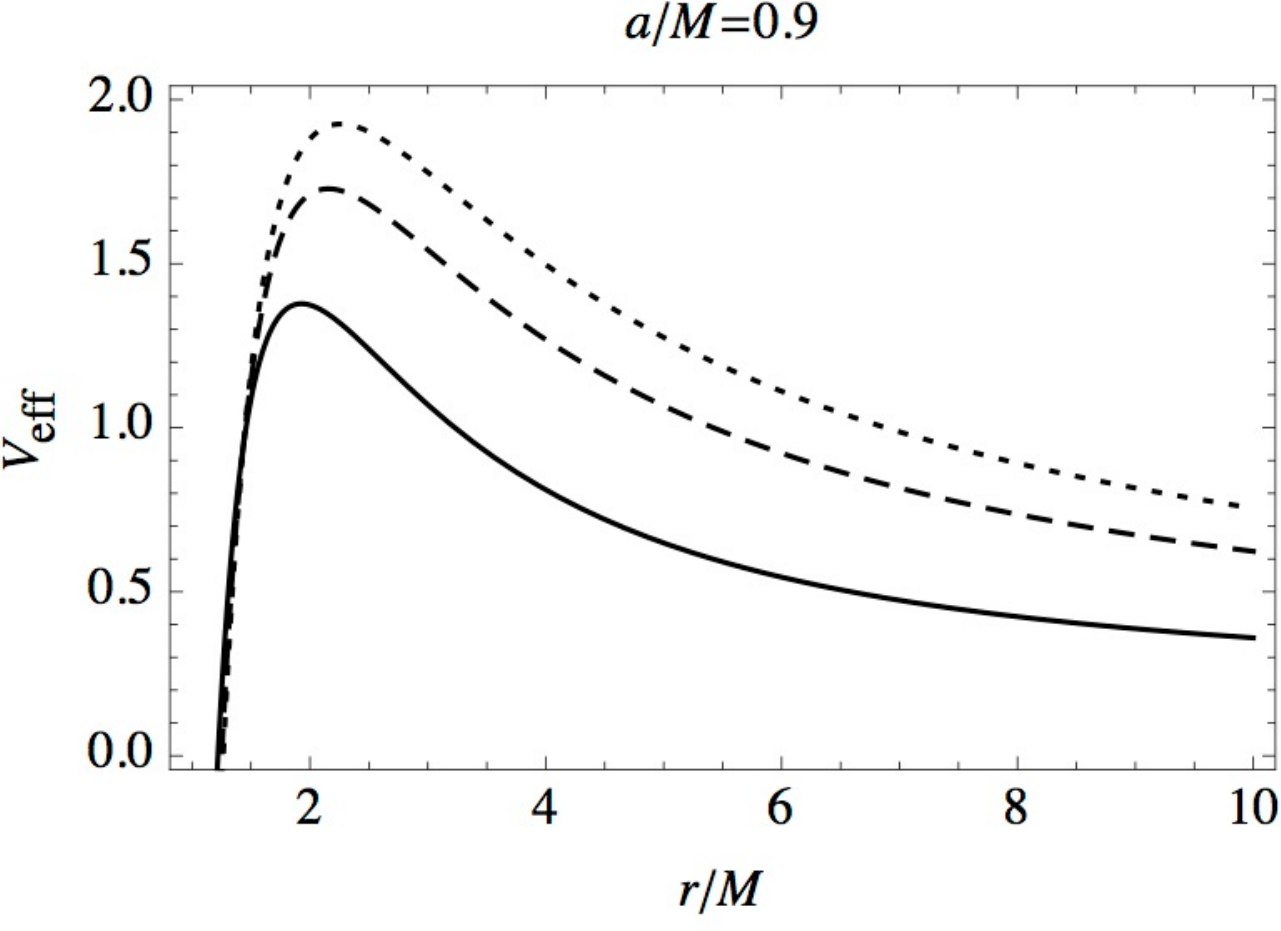}
\end{center}
\caption{ The radial dependence of the effective potential of radial motion of photons for the different values of rotation parameter and refraction index of the plasma.  Here the quantity $V_{\rm eff }$ is normalised by the energy of the photon ${\cal E}$.  \label{effpot}}
\end{figure*}

The radial dependence of the effective potential for different values of plasma refraction $n$ and black hole spin $a$ has been presented in Fig.~\ref{effpot}. In the Fig.\ref{effpot} the left plot corresponds to the case when refraction parameter of the plasma is $n^2= 0.2\ ,\ 0.44\ ,\ 0.89$  (dotted, dashed and solid lines, respectively) at the position $r=3M$; middle plot corresponds to the case when the refraction parameter is  $n^2= 0.19\ ,\ 0.42\ ,\ 0.88$  corresponding to dotted, dashed and solid lines, respectively, at the position $r=3M$; right plot represents the radial dependence of the effective potential when refraction parameter is $n^2= 0.14\ ,\ 0.39\ ,\ 0.88$  corresponding to dotted, dashed and solid lines, respectively, at the position $r=3M$.

\section{Shadow of black hole in the presence of the plasma }
\label{bh-shadow}
\begin{figure*}[t!]
\begin{center}
\includegraphics[width=0.222\linewidth]{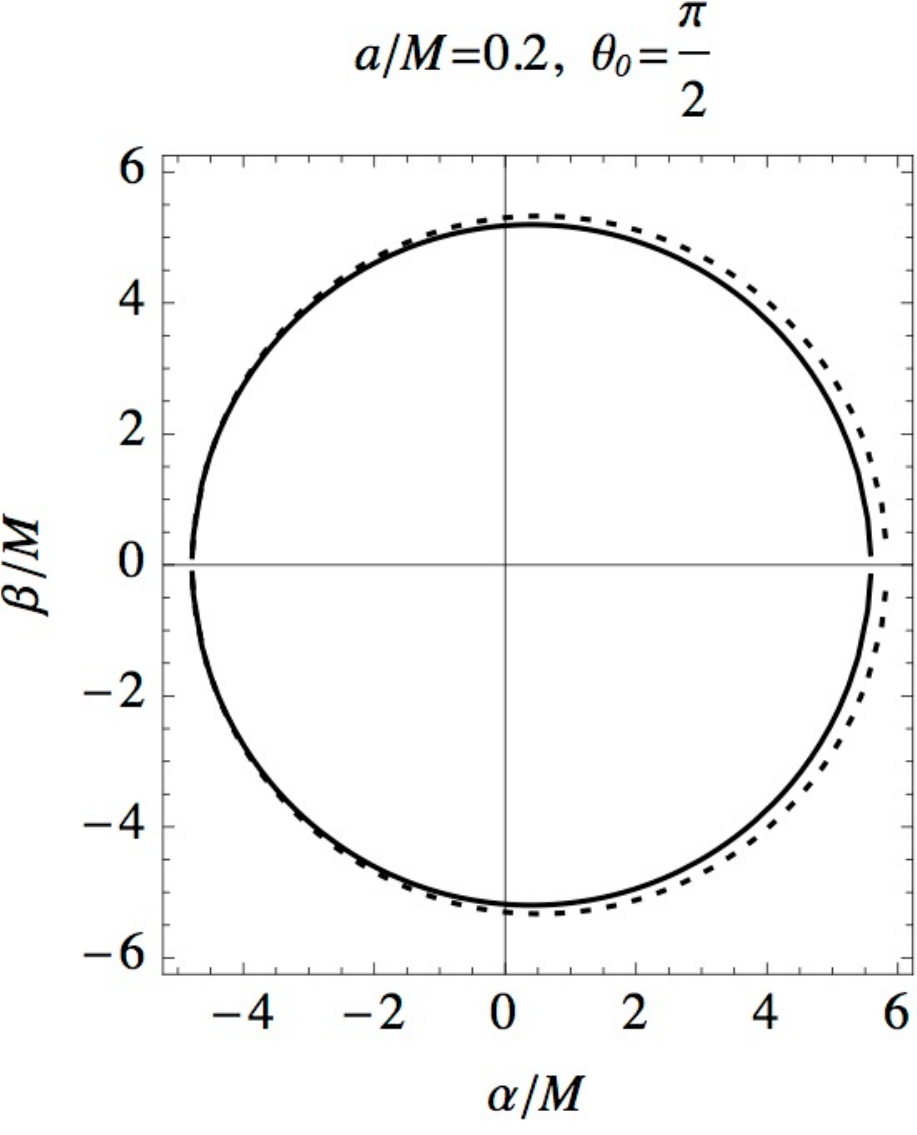}
\includegraphics[width=0.24\linewidth]{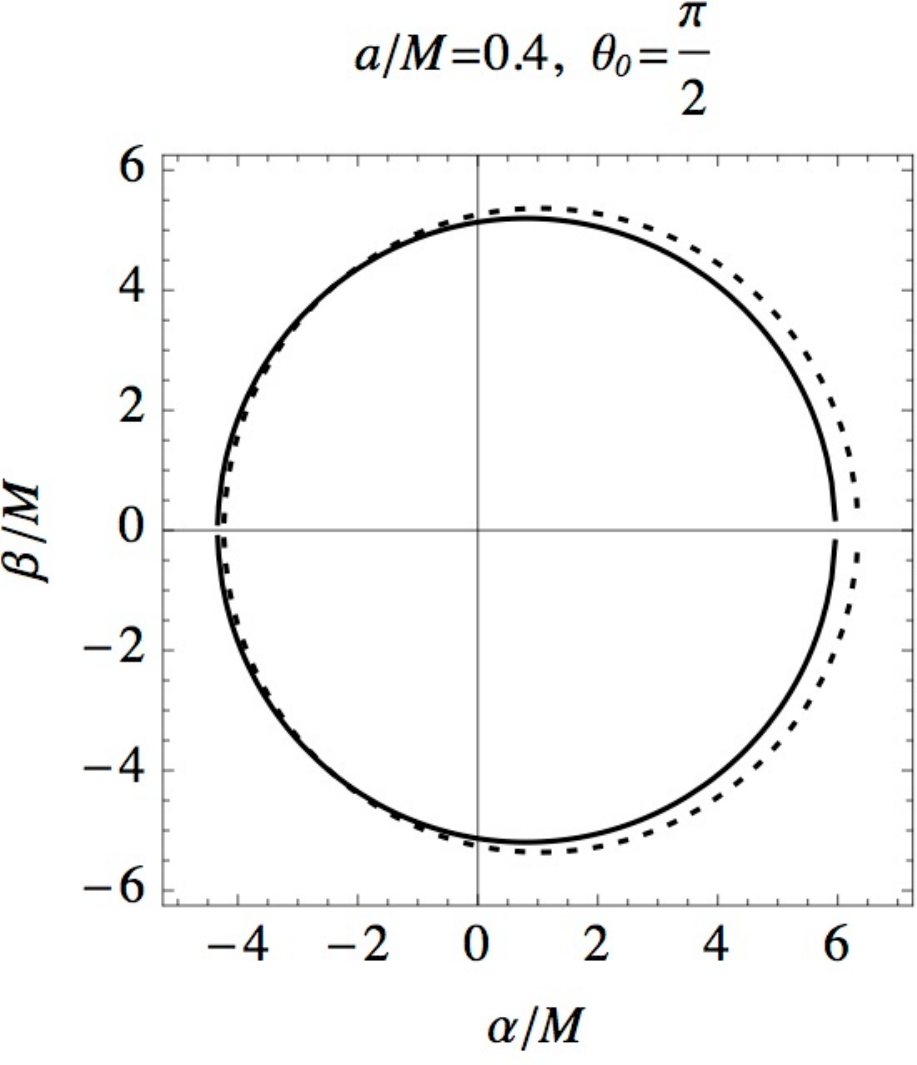}
\includegraphics[width=0.24\linewidth]{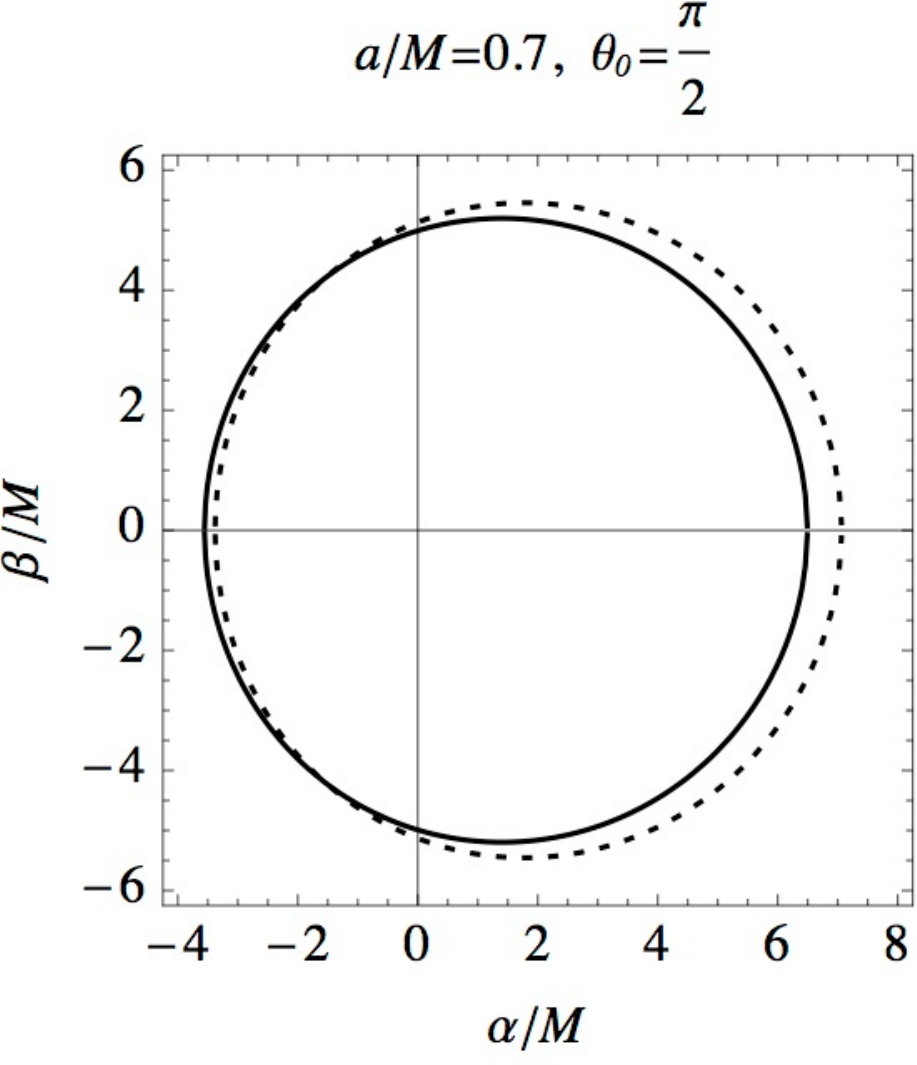}
\includegraphics[width=0.24\linewidth]{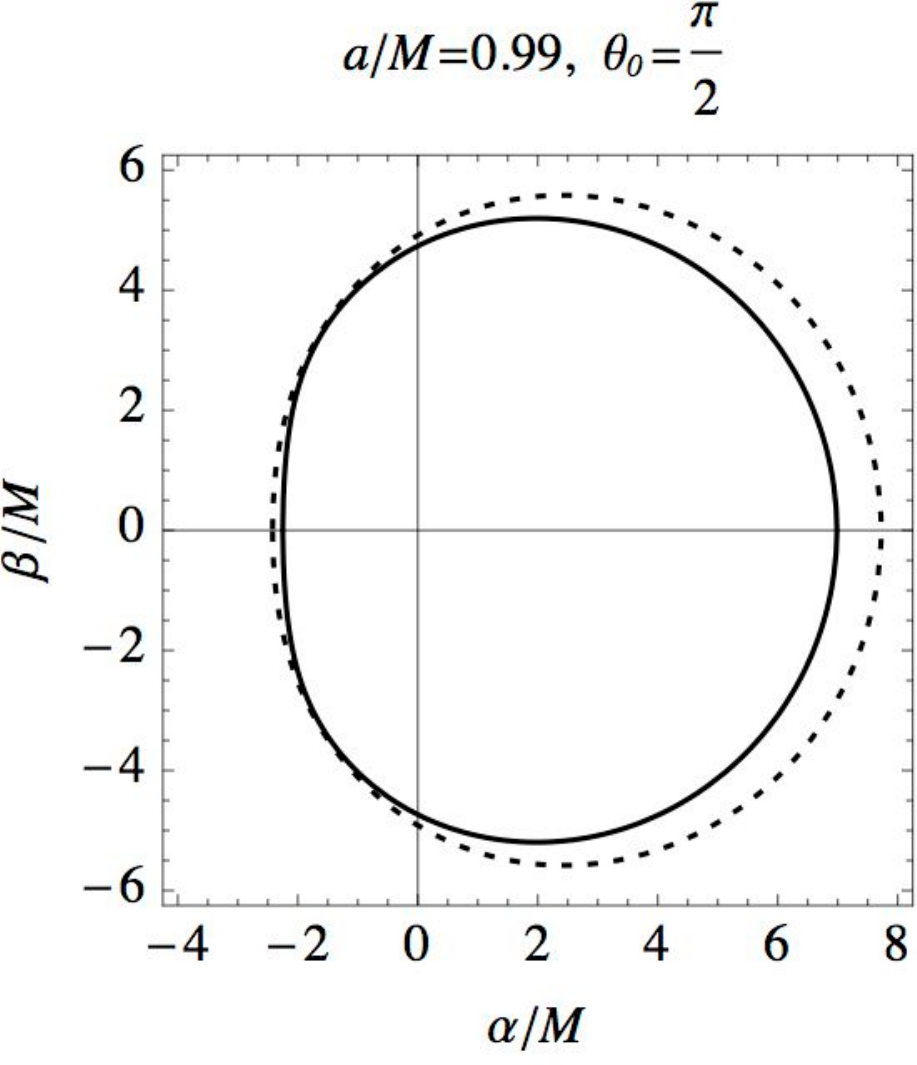}

\vspace{.4cm}
\includegraphics[width=0.222\linewidth]{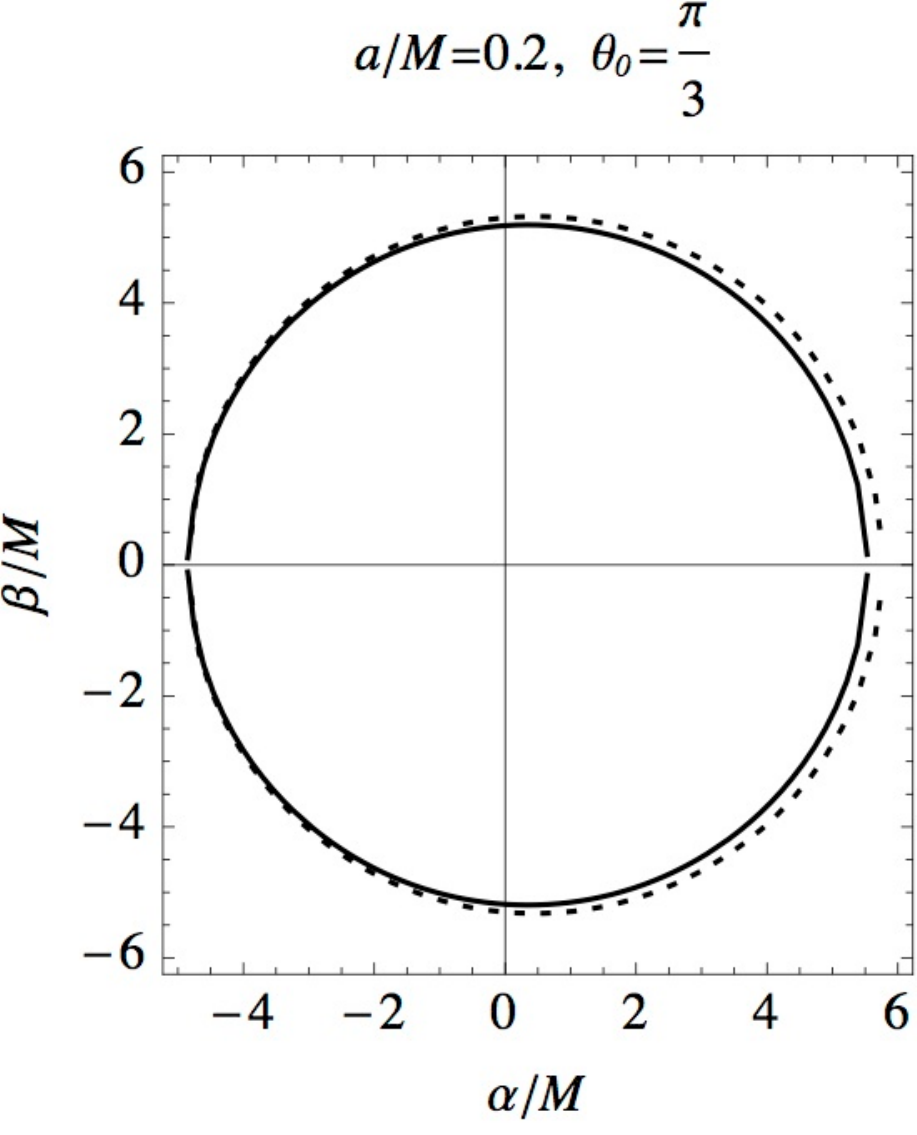}
\includegraphics[width=0.24\linewidth]{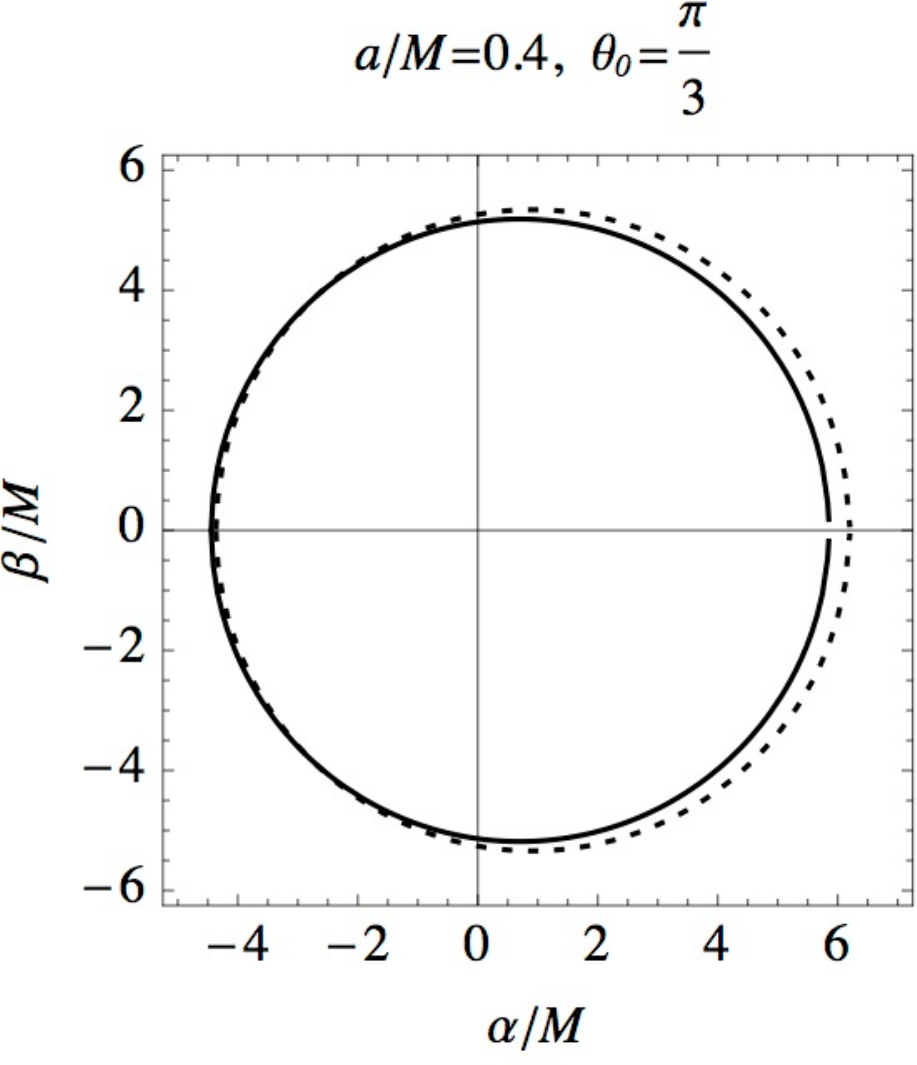}
\includegraphics[width=0.24\linewidth]{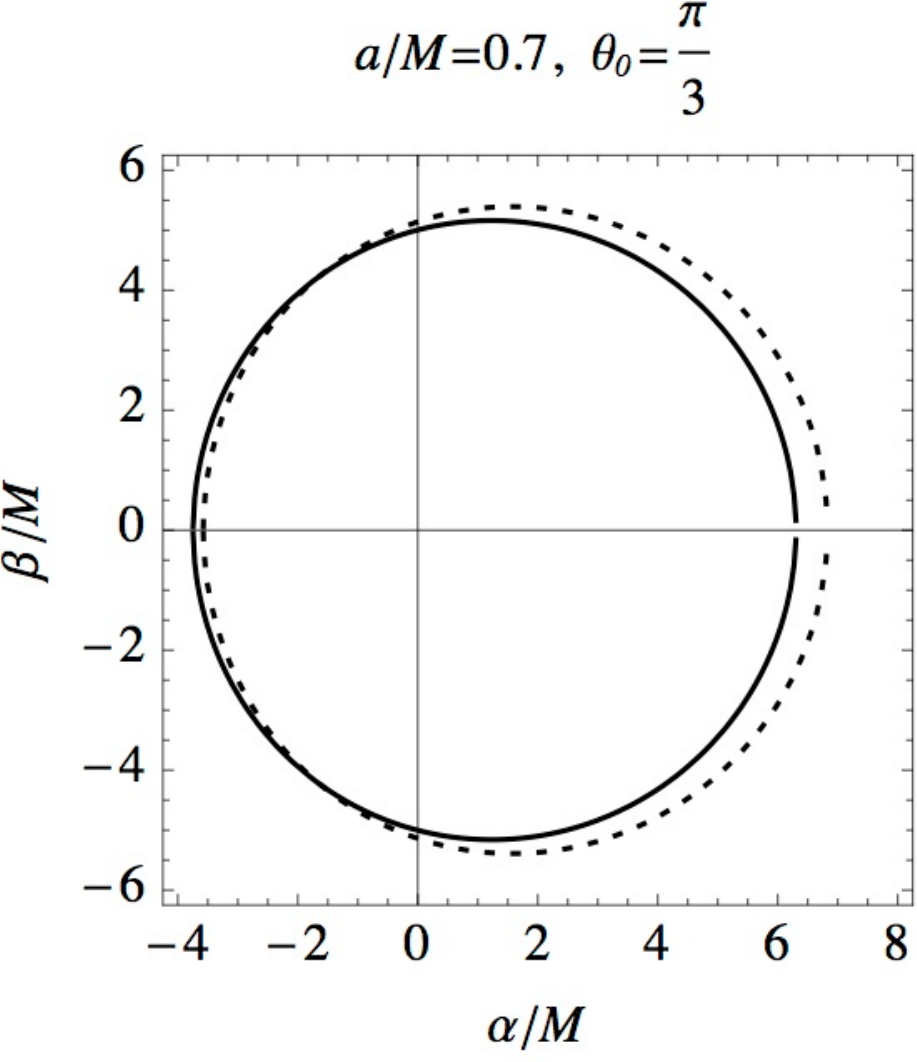}
\includegraphics[width=0.24\linewidth]{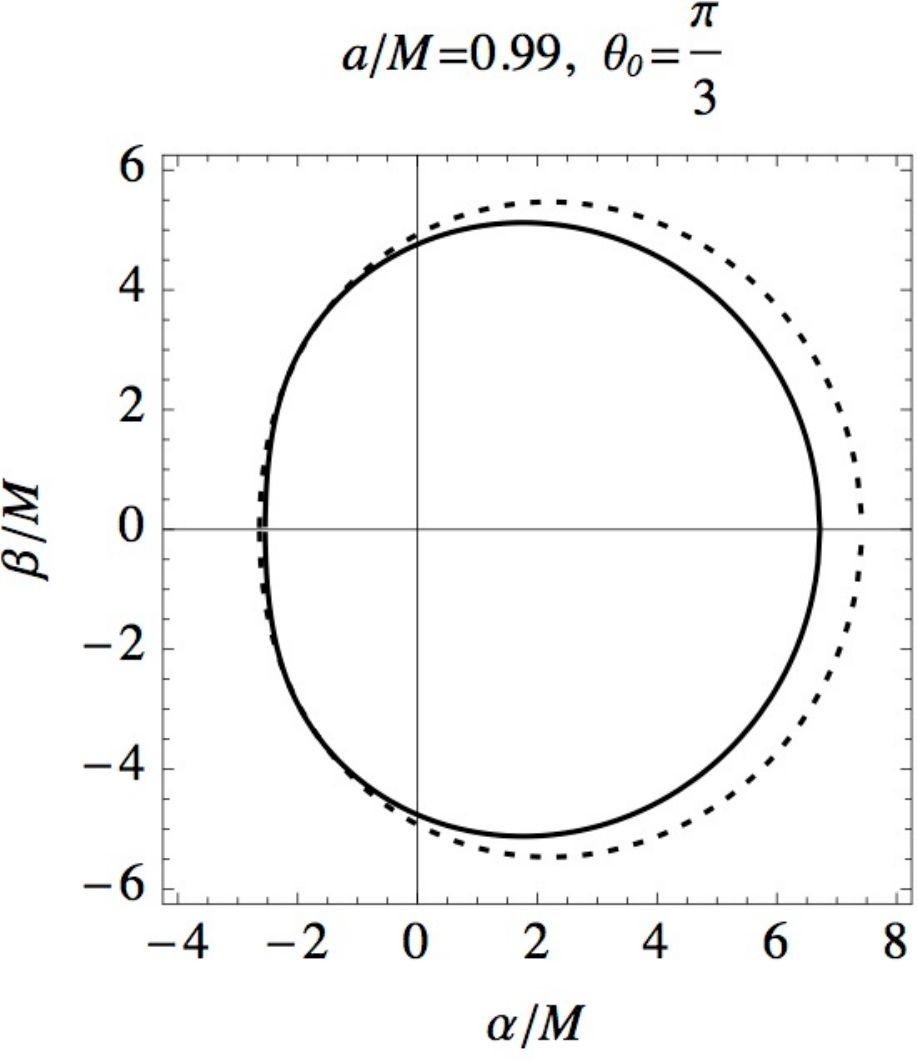}

\vspace{.4cm}
\includegraphics[width=0.222\linewidth]{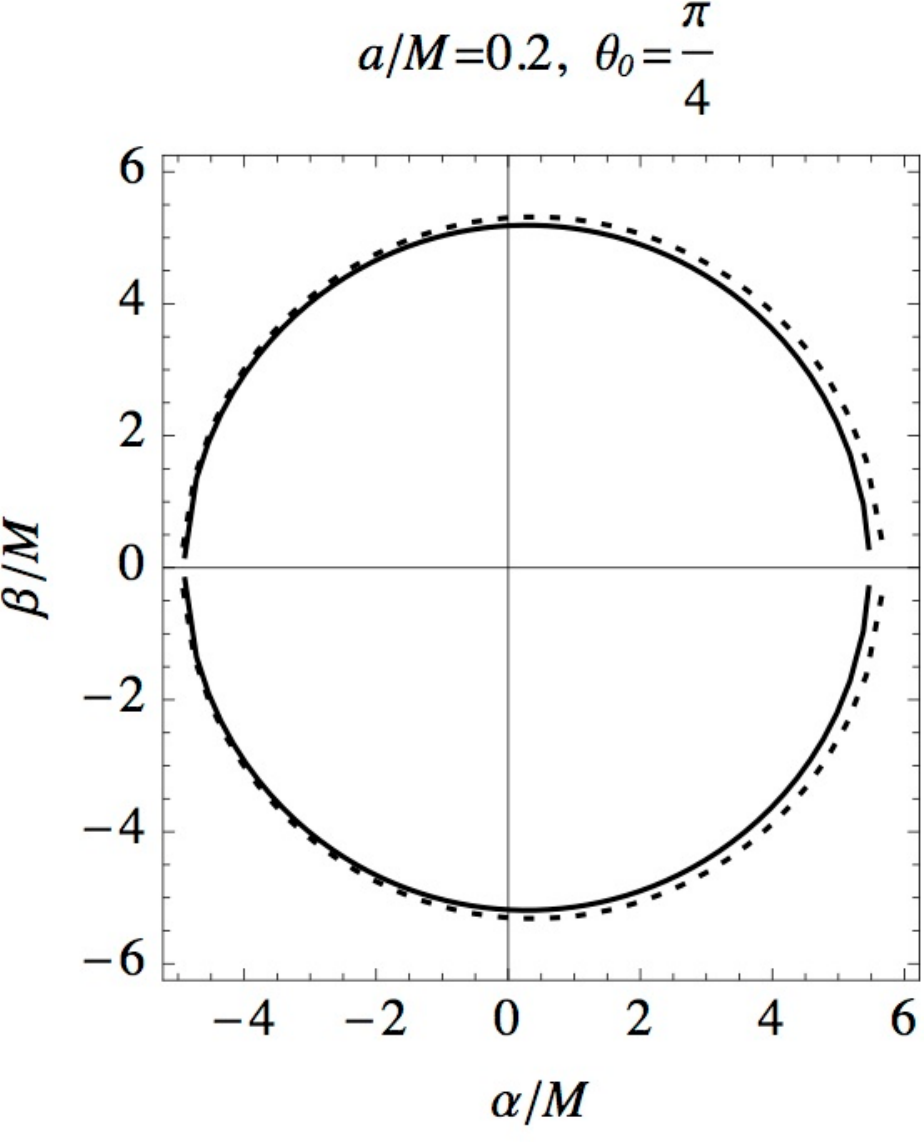}
\includegraphics[width=0.24\linewidth]{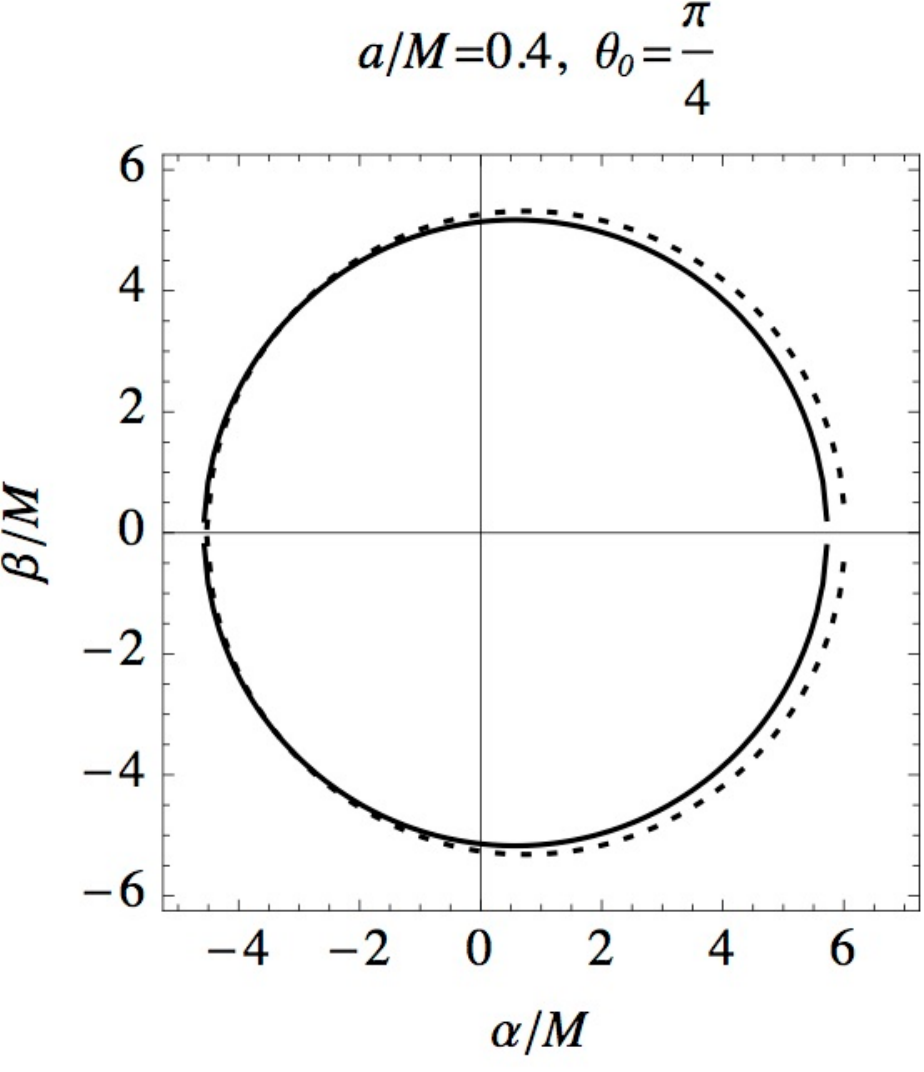}
\includegraphics[width=0.24\linewidth]{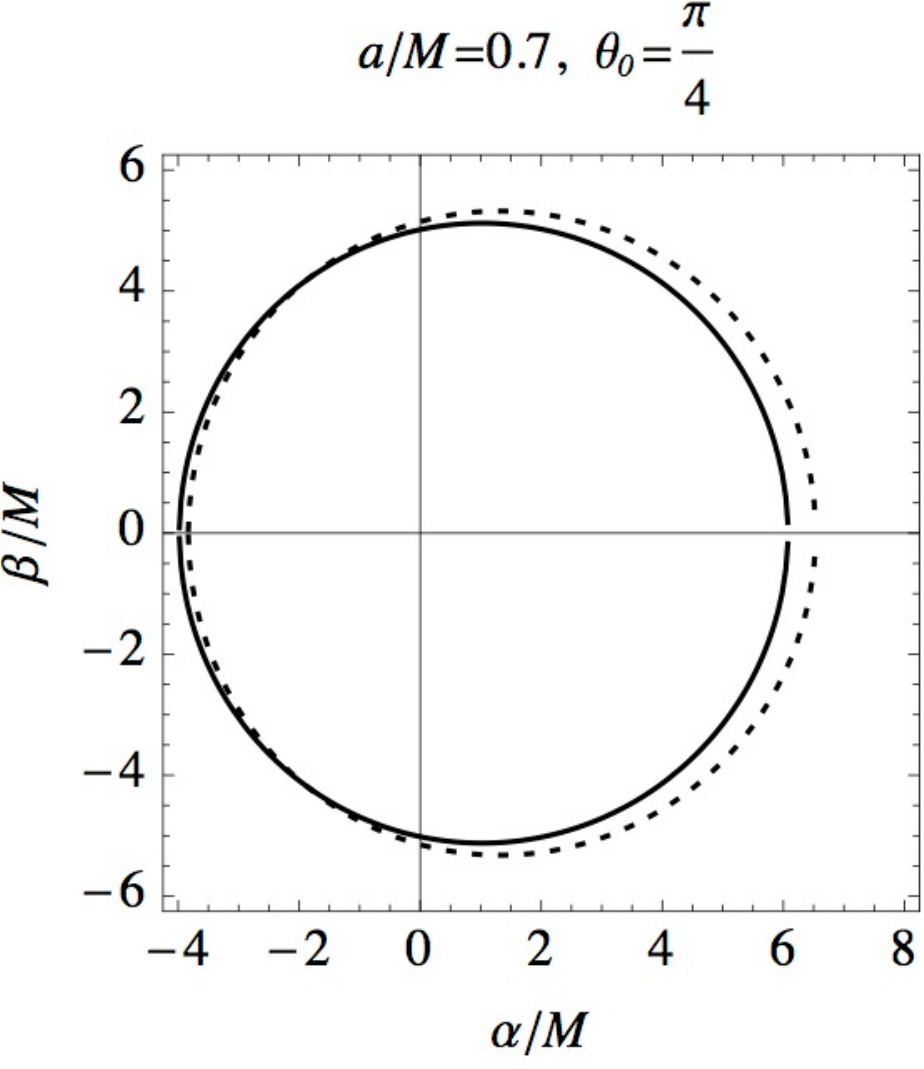}
\includegraphics[width=0.24\linewidth]{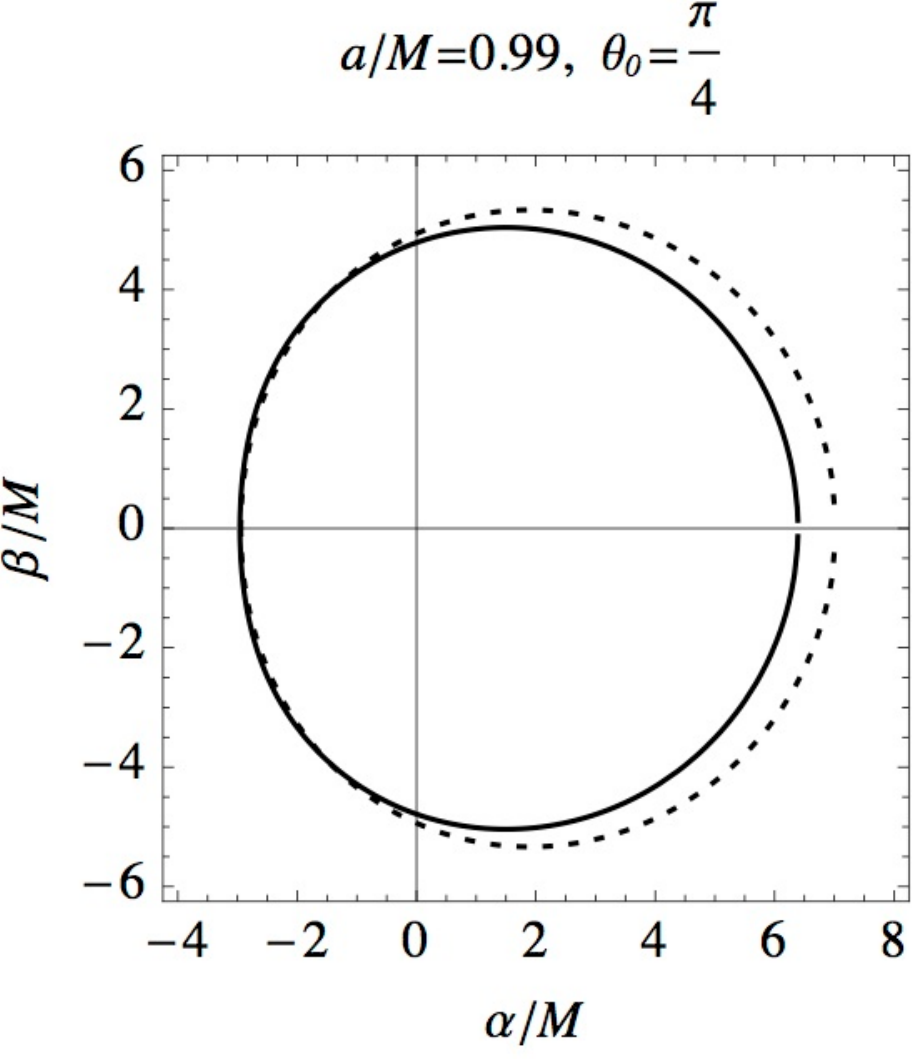}

\vspace{.4cm}
\includegraphics[width=0.24\linewidth]{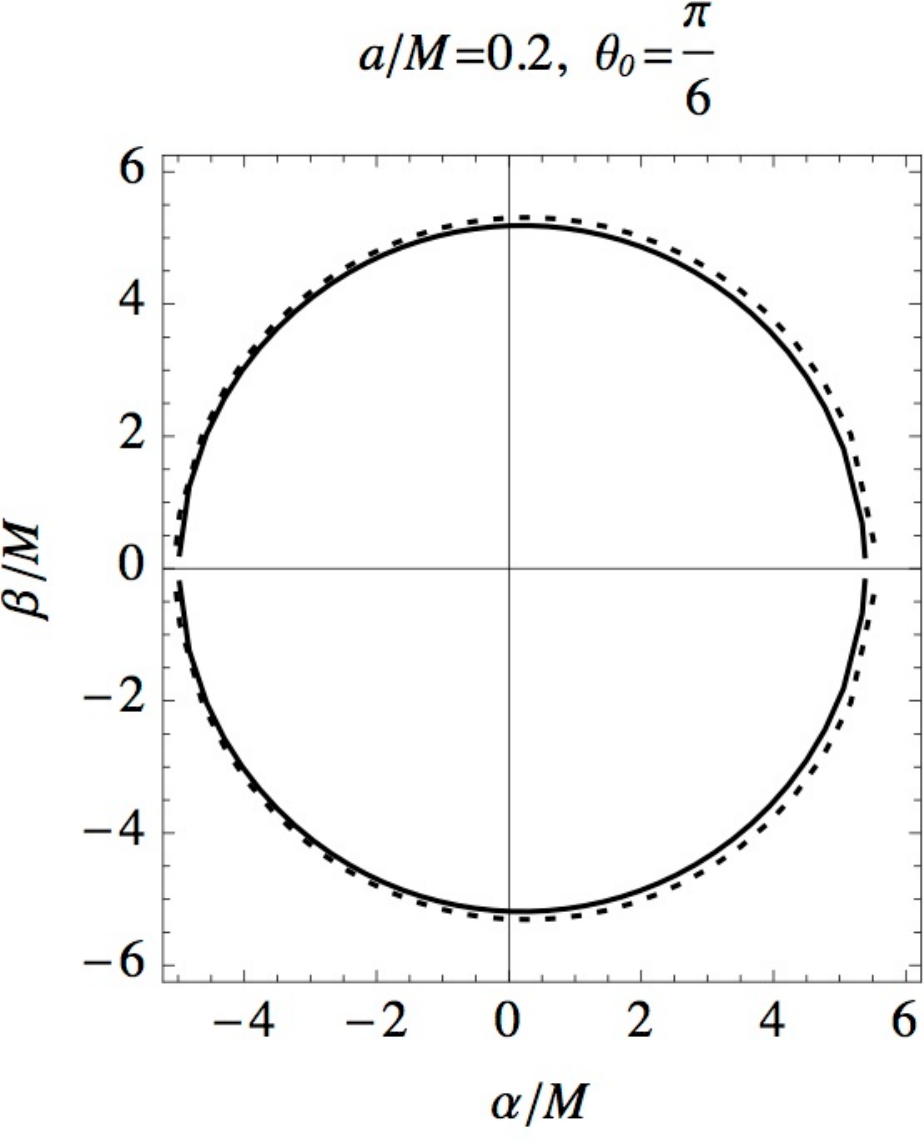}
\includegraphics[width=0.24\linewidth]{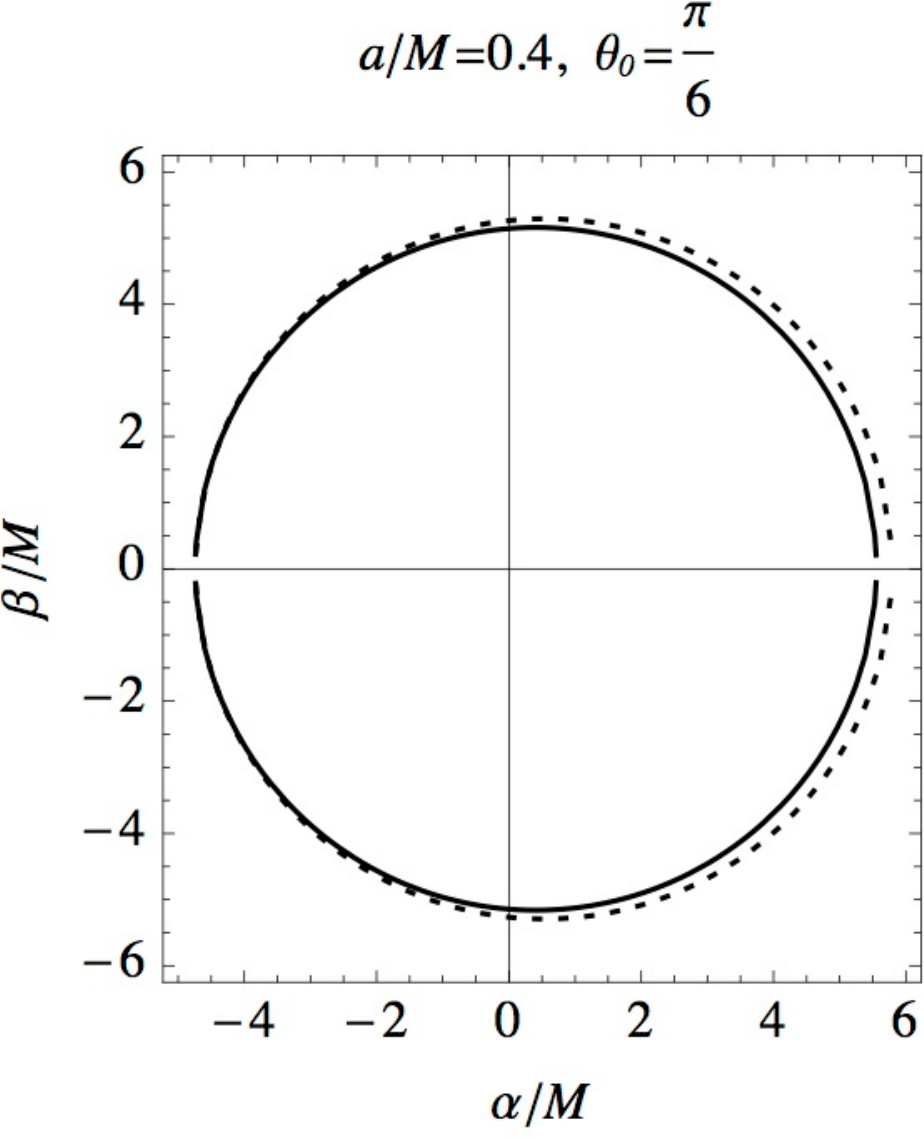}
\includegraphics[width=0.258\linewidth]{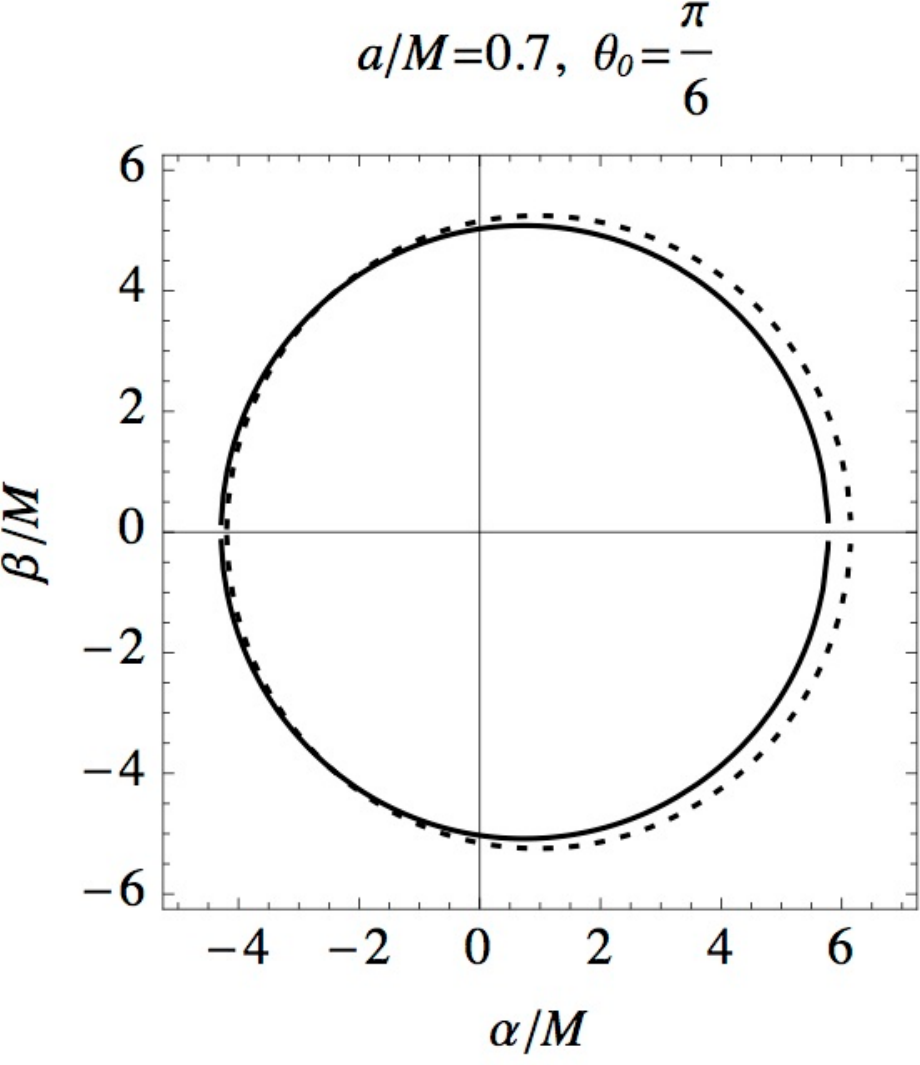}
\includegraphics[width=0.24\linewidth]{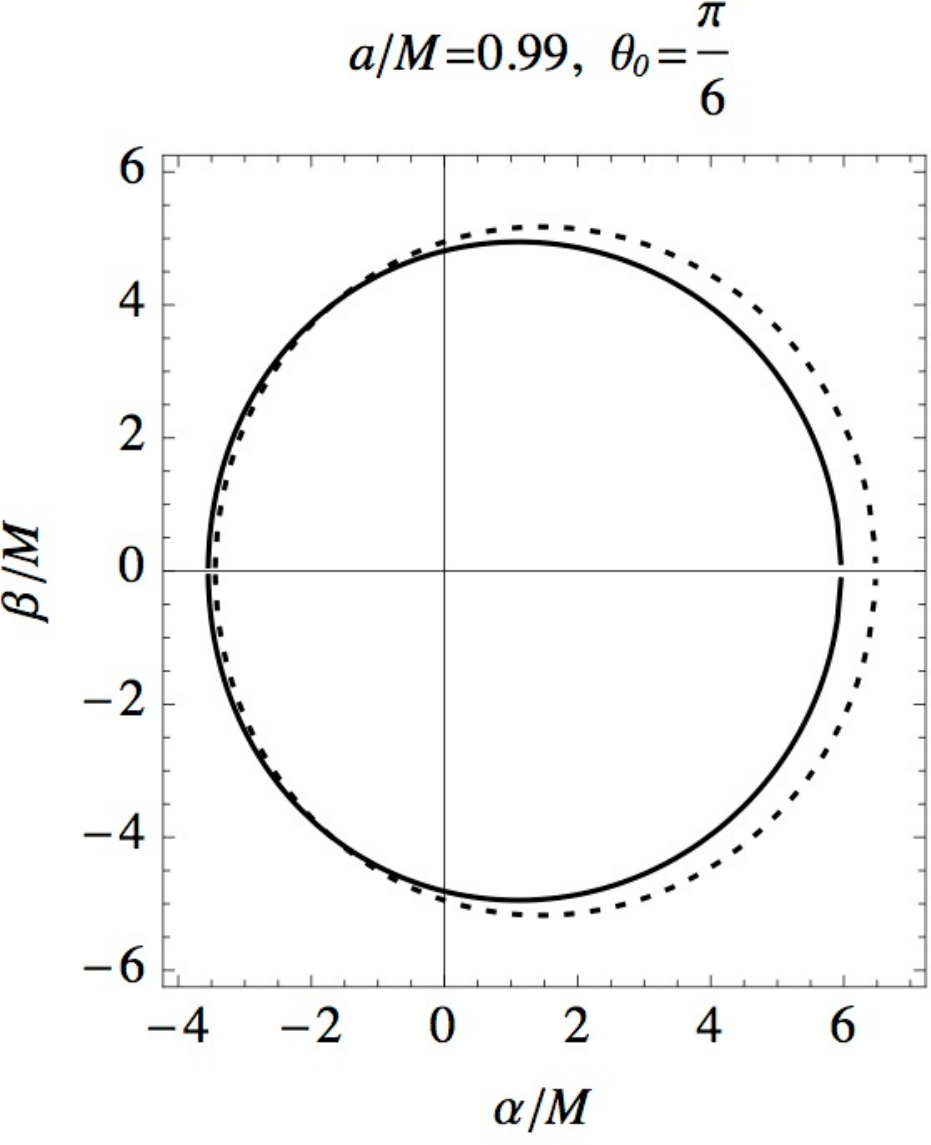}
\end{center}
\caption{The shadow of the black hole surrounded by plasma for the different values of the rotation parameter $a$, inclination ante between observer and the axis of the rotation $\theta_0$, and the refraction index. The solid lines in the plots correspond to the vacuum case, while for dashed lines we choose the plasma frequency $\omega_e/\omega_\xi=k/r$ and $(k/M)^2 = 0.5$\ . \label{shadow1}}
\end{figure*}

In this section we consider the shadow cast by black hole surrounded by plasma.
If black hole surrounded by plasma originated between the light source and the observer, then the latter can observe the black spot on the bright background. The observer at the infinity can only observe the light beam scattered away and due to capturing of the photons by the black hole the shaded area on the sky would be appeared. This spot corresponds to the shadow of the black hole and its boundary can be defined using the equation of motion of photons given by expressions (\ref{teqn})--(\ref{theteqn}) around black hole surrounded by plasma.

In order to describe the apparent shape of the the black hole surrounded by plasma we need to consider the closed orbits around it. Since the equations of motion depend on conserved quantities ${\cal E}$, ${\cal L}$ and the Carter constant ${\cal K}$, it is convenient to parametrize them using the normalised parameters
$
\xi={\cal L/E}
$ and $\eta= {\cal K/E}^2$.
The silhouette of the black hole shadow in the presence of the plasma can be found using the conditions
\[{\cal R}(r)=0=\partial {\cal R}(r)/\partial r . \]
Using these equations one can easily find the expressions for the parameters $\xi$ and $\eta$ in the form
\begin{eqnarray}
\xi &=&  \frac{\cal {B}}{{\cal A}} +\sqrt{\frac{{\cal B}^2}{{\cal A}^2} -\frac{{\cal C}}{{\cal A}}}\ , \label{xiexp}\\
\eta&=& \frac{(r^2+a^2-a\xi)^2 +(r^2+a^2)^2 (n^2-1)}{\Delta} \nonumber\\&& -(\xi-a)^2 \label{etaexp}
\end{eqnarray}
where we have used the following notations
\begin{eqnarray}
{\cal A}&=& \frac{a^2}{\Delta} \ ,\\
{\cal B}&=& \frac{a^2-r^2}{M-r}\frac{Ma }{\Delta} \ ,\\
{\cal C}&=& n^2 \frac{(r^2+a^2)^2}{\Delta}\nonumber \\&&+\frac{2r (r^2+a^2)n^2 +(r^2+a^2)^2 n n'}{M-r} \ ,
\end{eqnarray}
and prime denotes the differentiation with respect to radial coordinate $r$.

The boundary of the black hole's shadow can be fully determined through the expessions (\ref{xiexp})-(\ref{etaexp}). However, the shadow will be observed at 'observer's sky', which can be referenced by the celestial coordinates related to the real astronomical measurements. The celestial coordinates are defined as
\begin{eqnarray}  \label{alpha1}
\alpha&=&\lim_{r_{0}\rightarrow \infty}\left(
-r_{0}^{2}\sin\theta_{0}\frac{d\phi}{dr}\right)\ , \\
 \label{beta1}
\beta&=&\lim_{r_{0}\rightarrow \infty}r_{0}^{2}\frac{d\theta}{dr}\ .
\end{eqnarray}
Using the equations of motion (\ref{teqn})-(\ref{theteqn}) one can easily find the relations for the celestial coordinates in the form
\begin{eqnarray}
\alpha&=& -\frac{\xi}{n\sin\theta}\, \label{alpha}\ ,\\
\beta&=&\frac{\sqrt{\eta+a^2-n^2a^2\sin^2\theta-\xi^2\cot^2\theta }}{n} \label{beta}\ ,
\end{eqnarray}
for the case when black hole is surrounded by plasma.

In Fig~\ref{shadow1} the shadow of the rotating  black hole for the different values of black hole rotation parameter $a$, inclination angle $\theta_0$ between the observer and the axis of the rotation is represented. In this figures we choose the plasma frequency in the form $\omega_e/\omega_\xi=k/r$. From the Fig.~\ref{shadow1} one can observe the change of the size and shape of the rotating black hole surrounded by plasma. Physical reason for this is due to gravitational redshift of photons in the gravitational field of the black hole. The frequency change due to gravitational redshift affects on the plasma refraction index.
\subsection{\label{nonrot}Shadow of non-rotating black hole}

\begin{figure}[t!]
\includegraphics[width=0.9\linewidth]{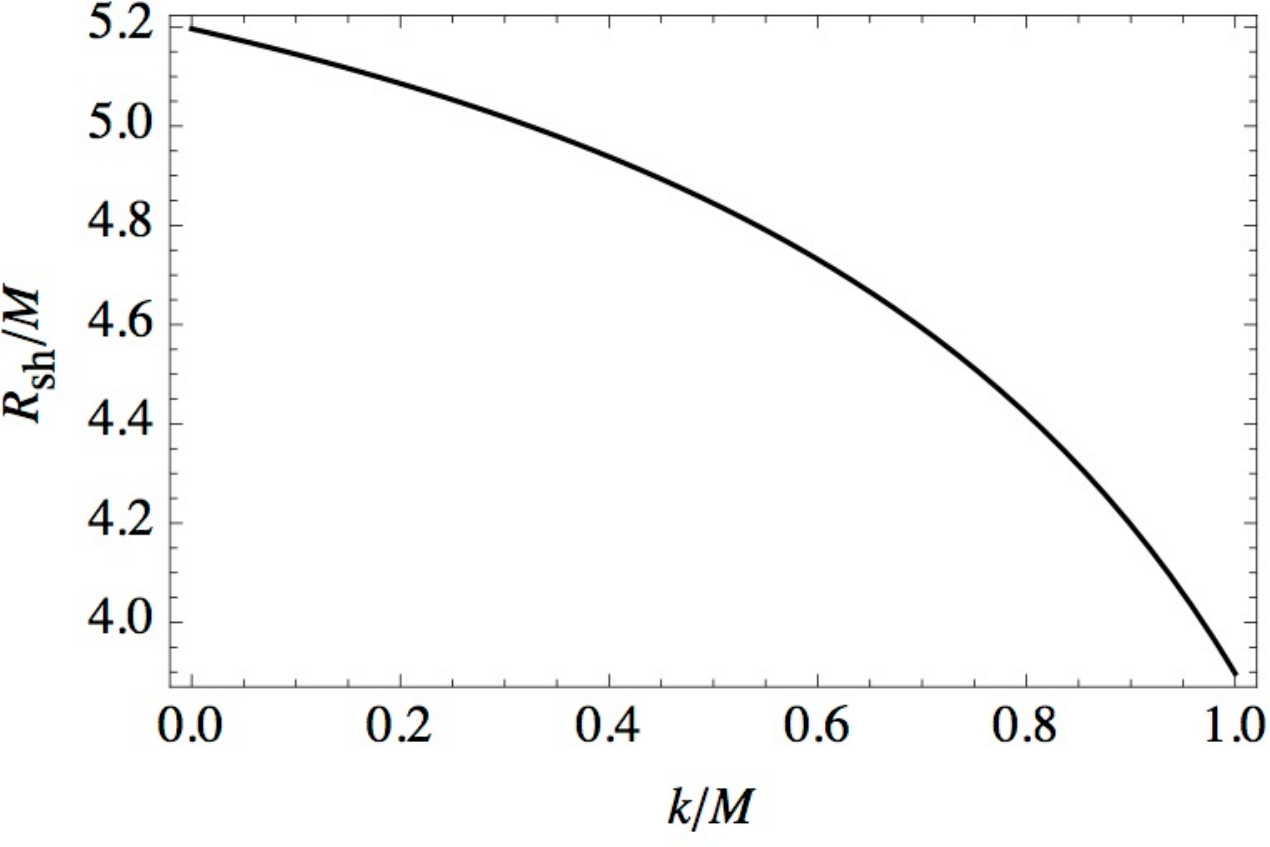}
\caption{ The dependence of the Schwarzschild black hole shadow on the plasma frequentcey parameter. Here we take the plasma frequency as $\omega_e/\omega_\xi=k/r$.  \label{fig3}}
\end{figure}

Now in order to extract pure plasma effects we will concentrate at the special case when the black hole is non-rotating and the size of the black hole shadow can be observed (see, e.g.  \cite{Perlick15}). In the case of the static black hole the shape of the black hole is circle and the radius of the shadow will be changed by plasma effects. Using the expressions (\ref{alpha}) and (\ref{beta}) one can easily find the radius of the shadow of static black hole surrounded by plasma in the form:
\begin{eqnarray}
R_{sh}&=& \frac{1}{n(r-M)}\Bigg[2 r^3 (r-M) n^2   +r^4 nn'(r-M)\nonumber\\&&-2 r^2 M^2   + 2M r^2  \bigg\{n r^2 (n+r n') - (4 n +3r n') \nonumber\\&&\times n Mr+M^2 (1+3 n^2+2r nn')\bigg\}^{1/2}\Bigg]^{1/2}\ ,\label{shadrad}
\end{eqnarray}
here $r$ is the unstable circular orbits of photons defined by $dr/d\sigma = 0$ and $\partial V_{\rm eff}/\partial r =0$. In the absence of the plasma one has the standard value of the photon sphere radius as $r=3M$ and for shadow radius $R_{sh}=3\sqrt{3}$~\cite{Virbhadra00,Claudel01}. In the presence of the plasma we will have different value for the photon sphere radius and consequently different shadow radius of the boundary of black hole shadow. In the Fig.~\ref{fig3} the dependence of the radius of shadow of the static black hole from the plasma parameters has been presented which shows that  the radius of the shadow of black hole surrounded by inhomogeneous plasma decreases. It is similar to 
the results of the paper~\cite{Perlick15}.

\subsection{Emission energy of black holes in plasma \label{emission}}

\begin{figure}[t!]
\includegraphics[width=0.9\linewidth]{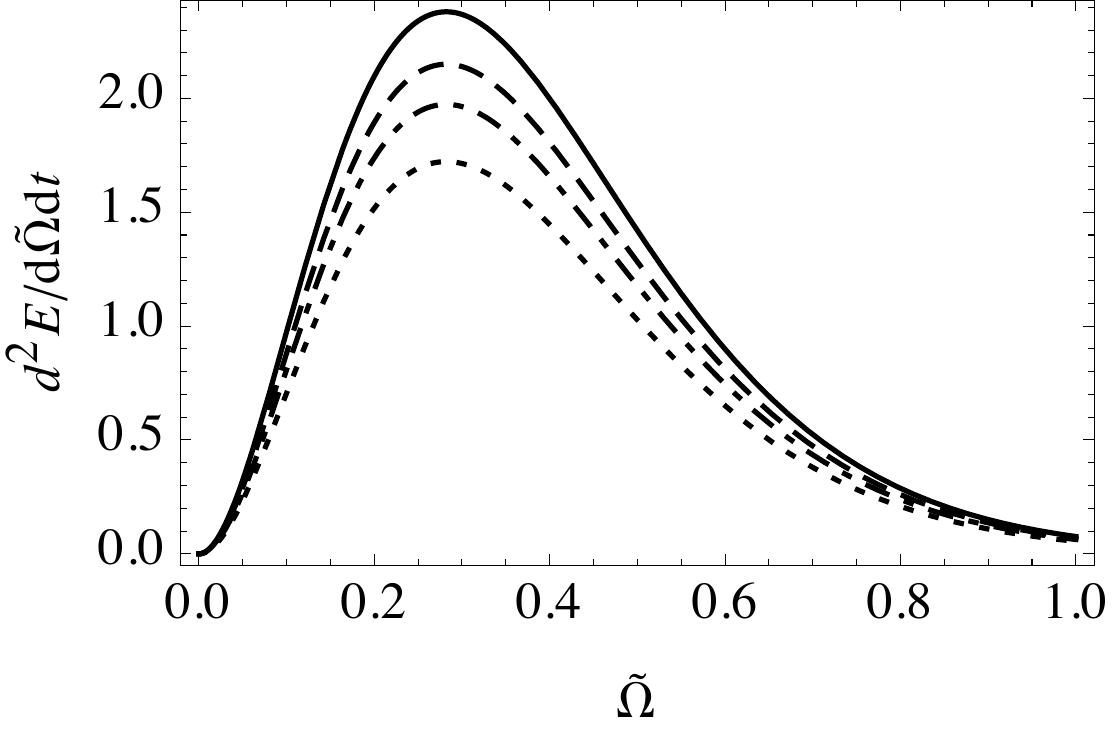}
\caption{  Energy emission from black hole  for the different values of $k/M$: Solid line corresponds to the vacuum case ($k/M=0$), Dashed line corresponds to the case when $k/M=0.4$, dot-dashed line corresponds to the case when $k/M=0.6$, dotted line corresponds to the case when $k/M=0.8$. Here we take the plasma frequency as $\omega_e/\omega_\xi=k/r$ and $\tilde{\Omega}$ is normalised to the $T$.  \label{energy}}
\end{figure}

For the completeness of our study here we evaluate rate of the energy emission from the
black hole in plasma using the expression for the Hawking radiation at
the frequency $\Omega$ as~\cite{Wei,Atamurotov15}
\begin{equation}
\frac{d^2E(\Omega)}{d\Omega dt}= \frac{2 \pi^2 \sigma_{lim}}{\exp{\Omega/T}-1}\Omega^3\ ,
\end{equation}
where $T=\kappa/2\pi$ is the Hawking temperature, and $\kappa$ is the surface gravity. Here, for simplicity we consider the special case when the black hole is non rotating, and the background spacetime is spherically-symmetric.

At the horizon the temperature $T$ of the black hole is
%\nonumber \\ &&
\begin{eqnarray}
T&=& \frac{1}{4\pi r_{+}}\ .
\label{temp}
\end{eqnarray}

The limiting constant $\sigma_{lim}$
\begin{equation}
\sigma_{lim} \approx \pi R_{sh}^2\  \nonumber
\end{equation}
defines the value of the
absorption cross section vibration for spherically
symmetric black hole and $R_{sh}$ is given by expression (\ref{shadrad}).

Consequently,  one can get
\begin{eqnarray}
\frac{d^2E(\Omega)}{d\Omega dt}=\frac{2\pi^3 R_{sh}^2}{e^{\Omega/T}-1}\Omega^3\  \nonumber
\end{eqnarray}
that the energy of radiation of black hole in plasma depends on the size of its shadow.

{The dependence of energy emission rate from frequency for the different values of plasma parameters} $\omega_{e}$ is shown in Fig.~\ref{energy}.
{One can see that with the increasing plasma parameter} $\omega_{e}$  {the maximum value of energy emission rate decreases, caused by radius of shadow  decrease.}

\section{Conclusions}
\label{conclusion}
In this paper we have studied shadow and emission rate of axial symmetric black hole in presence of plasma with radial power-law density.
The obtained results can be summarized as follows.

\begin{itemize}

\item In the presence of plasma the observed shape and size of shadow changes depending on i) plasma parameters, ii) black hole spin
and iii) inclination angle between observer plane and axis of rotation of black hole.

\item In order to extract pure effect of plasma influence on
black hole image the particular case of the Schwarzschild black hole has also been investigated.
It is shown that under
influence of plasma the observed size of shadow of the spherical symmetric black hole becomes smaller than that in the vacuum case.
So it has been shown
that i) the photon sphere around the spherical symmetric black hole is left unchanged under the plasma influence, ii) however
the Schwarzschild black hole shadow size in plasma is reduced due to the refraction of the electromagnetic radiation
in plasma environment of black hole.

\item The study of the energy emission from the black hole in plasma has shown that with the increase of the dimensionless
plasma parameter the maximum value of energy emission rate from the black hole decreases due to the decrease of the size of black hole
shadow.

\end{itemize}

In the future work we plan to study shadow and related optical properties  of different types of gravitational compact objects in the presence of plasma in more detail and in
more astrophysically relevant cases.

%%%%%%%%%%%%%%%%%%%%%%%%%%%%%%%%%%%%%%%%%%%%%%%%%%%%%%%%%%%%%%%%%%%%%

%\acknowledgments

%%%%%%%%%%%%%%%%%%%%%%%%%%%%%%%%%%%%%%%%%%%%%%%%%%%%%%%%%%%%%%%%%%%%%%%%%
\section*{Acknowledgments}
%%%%%%%%%%%%%%%%%%%%%%%%%%%%%%%%%%%%%%%%%%%%%%%%%%%%%%%%%%%%%%%%%%%%%%%%%

This research
was partially supported by the Volkswagen Stiftung (Grant
86 866), by the project F2-FA-F113 of the UzAS
and by the ICTP through the projects OEA-NET-76, OEA-PRJ-29.
Warm hospitality
that  has facilitated this work to A.A. and B.A.  by the Goethe University, Frankfurt
am Main, Germany and the IUCAA, Pune is thankfully acknowledged. A.A. and B.A. acknowledge the TWAS
associateship grant. We would like to thank Volker Perlick and anonymous referee for the careful reading
and important comments which essentially improved the paper.

\bibliographystyle{apsrev4-1}  %% BibTeX style
\bibliography{/hp/ahmadjon_hp/Nauka/gravreferences/gravreferences}

\end{document}